\documentclass[epj]{svjour}
\usepackage{epsfig}
\usepackage{graphics}
\begin{document}
\title{Sensitivity of Pion versus Parton-Jet Nuclear Modification Factors to the Path-Length Dependence of Jet-Energy Loss at RHIC and LHC}
\author{Barbara Betz\inst{1} \and Miklos Gyulassy\inst{2}}

\titlerunning{Sensitivity of Pion vs.\ Parton-Jet Nuclear Modification Factors
at RHIC and LHC}

\institute{Institut f\"ur Theoretische Physik, 
Johann Wolfgang Goethe-Universit\"at, Frankfurt am Main, Germany
\and 
Department of Physics, Columbia University, 
New York, 10027, USA}
\date{Received: date / Revised version: date}
%
\abstract{
We compare the jet-path length and beam-energy dependence of the pion nuclear modification factor
and a parton-jet nuclear modification factor at RHIC and LHC. We contrast 
predictions based on a linear pQCD and a highly non-linear hybrid-AdS holographic model of jet-energy loss. 
We find that both models require a reduction of the jet-medium coupling from RHIC to LHC 
to account for the measured pion nuclear modification factor. In case of the parton-jet nuclear 
modification factor, however, which serves as a lower bound for the LO jet nuclear modification factor of reconstructed jets, 
the extracted data can be characterized without a reduced jet-medium 
coupling at LHC energies. We conclude that while reconstructed jets are sensitive to both quarks and gluons 
and thus provide more information than the pion nuclear modification factor, 
their information regarding the jet-medium coupling is limited due to the superimposition with
NLO and medium effects. Hence, a detailed description of the underlying physics 
requires both the leading hadron and the reconstructed jet nuclear modification factor.
Unfortunately, the results for both the pion and the parton-jet nuclear modification factor 
are insensitive to the jet-path dependence of the models considered. 
\PACS{
      {12.38.Mh}{Quark-gluon plasma}   \and
      {25.75.Bh}{Hard scatterings in relativistic heavy-ion collisions} \and
      {11.25.Tq} {Gauge/string duality}
     } 
} 

\maketitle
\section{Introduction}
Jet-quenching observables have proven to render important information about
the evolution of the quark-gluon plasma (QGP) created in heavy-ion collisions
\cite{JET,reviews}. It got evident that {\it both} the jet-medium dynamics and 
the bulk medium evolution influence the jet-quenching observables 
\cite{Adare:2012wg,Betz:2014cza,Xu:2014ica,WHDG11,Jia:2011pi,Renk:2011aa,Marquet:2009eq}.
However, it remains a formidable task to identify the details of the 
jet-medium interactions and the jet-energy loss formalism. 

While it was shown that the "surprising transparency" of the medium
created at the LHC \cite{Xu:2014ica,WHDG11,Betz:2012qq,Buzzatti:2012dy} can
be described by perturbative QCD (pQCD) processes including a running of the 
jet-medium coupling
\cite{Betz:2014cza,Xu:2014ica,WHDG11,Jia:2011pi,Renk:2011aa,Marquet:2009eq,Betz:2012qq,Buzzatti:2012dy,glv,formalisms,Zakharov:2012fp,Molnar:2013eqa,Liao:2008dk,DGLV}, 
it remains an open question if pQCD is the correct prescriptions for the jet-energy loss 
at RHIC and LHC or if string theory based on the AdS/CFT correspondence 
\cite{Casalderrey-Solana:2014bpa,Chesler:2014jva,Ficnar,Gubser:2008as} 
with strings shooting up or falling into black holes must be applied. 

A main difference between pQCD and AdS/CFT calculations of the jet-energy loss
is the path-length dependence \cite{Adare:2012wg,Betz:2014cza,Jia:2011pi}. 
Commonly, it is assumed that there is a linear path-length dependence 
($dE/dx\sim x^{z=1}$) for pQCD processes and a squared path-length dependence 
($dE/dx\sim x^{z=2}$) for AdS/CFT-based scenarios.

Recently, we performed a systematic study on the jet-energy loss \cite{Betz:2014cza}
based on a generic energy loss model \cite{Betz:2014cza,Betz:2012qq} that can interpolate 
between a pQCD and an AdS/CFT jet-energy loss prescription. We demonstrated that we cannot constrain 
the path-length dependence narrower than $z=[0,2]$. In particular, we showed that the 
rapid rise of the pion nuclear modification factor $R_{AA}$ at LHC energies disfavors a 
{\it conformal} AdS/CFT prescription (with the same jet-medium coupling at RHIC and LHC).
{\it Non-conformal} AdS/CFT, however, with a jet-medium coupling reduced at 
LHC energies describes the measured pion data 
\cite{Adare:2012wg,Abelev:2012hxa,CMS:2012aa,CMS:2012rba}.

Last year, Casalderrey-Solana et al.\ introduced a "hybrid strong/weak approach" 
\cite{Casalderrey-Solana:2014bpa} based on the idea that any parton of a jet
propagating through a quark-gluon plasma suffers hard splitting, while additionally each
of these partons possesses soft fields that interact strongly with the medium.
The jet-energy loss prescription in Ref.\ \cite{Casalderrey-Solana:2014bpa} is based 
on strings falling into black holes featuring a non-linear Bragg peak 
\cite{Chesler:2014jva}. Casalderrey-Solana et al.\ showed that their ansatz 
reproduces the experimentally determined Jet $R_{AA}$ for reconstructed jets at LHC energies 
and renders a Jet $R_{AA}$ at RHIC energies that is very similar 
to the measured pion nuclear modification factor for {\it the 
same coupling} as considered for LHC energies, suggesting that the hybrid
strong/weak approach including the Bragg peak favors a conformal AdS/CFT 
prescription. In the following, we will refer to this energy-loss ansatz as Hybrid AdS.

In this paper, we determine the jet-energy loss based on radiative pQCD \cite{Betz:2014cza,Betz:2012qq}
and on the Hybrid AdS energy-loss ansatz of Ref.\ \cite{Casalderrey-Solana:2014bpa}. We
compare the pion nuclear modification factor with a parton-jet nuclear modification factor
that can be considered as an idealized LO Jet $R_{AA}$ at RHIC 
and LHC energies. For both scenarios, the measured pion nuclear modification 
factor can only be described if a reduced jet-medium coupling at LHC energies 
is considered. In case of the Jet $R_{AA}$, however, our results suggest as in 
Ref.\ \cite{Casalderrey-Solana:2014bpa} that the measured Jet $R_{AA}$ can 
possibly be characterized without a reduced jet-medium coupling at LHC energies. 

We conclude that the information of reconstructed jets regarding the jet-medium
coupling is limited and superimposed by NLO/medium effects. While reconstructed
jets do certainly provide more information regarding the quark and gluon contributions, 
the information on the jet-medium coupling cannot be extracted unambiguously. 
To do so, a leading hadron nuclear modification factor is needed. 
Besides that, the study shows that the Hybrid AdS energy-loss approach also favors
a {\it non-conformal} approach as the measured pion nuclear modification factor
can only be described with a reduced jet-medium coupling at LHC energies. 
We demonstrate that neither the pion nor the Jet 
$R_{AA}$ are sensitive to the path-length difference between the pQCD and the Hybrid AdS 
energy-loss model. Thus, we confirm that the path-length dependence 
for a jet-energy loss in heavy-ion collisions cannot be constrained further than 
$z=[0,2]$ \cite{Betz:2014cza}.

\section{The jet-energy loss models}
\label{Section2}

\subsection{The generic energy-loss model}
The jet-energy loss prescription for the pQCD process considered is based on our
generic jet-energy loss model \cite{Betz:2014cza,Betz:2012qq} that parametrizes the energy
loss via
\begin{eqnarray}
\frac{dE}{dx}=\frac{dE}{d\tau}= 
-\kappa(T)  E^a(\tau) \, \tau^{z} \, T^{c=2+z-a} \, \zeta_q
\;.
\label{Eq_pQCD}
\end{eqnarray}
Here, the jet-energy dependence, the path-length dependence, the temperature 
dependence, and the jet-energy loss fluctuations are characterized by the exponents 
$(a,z,$ $c,q)$. The jet-energy loss fluctuations are distributed via
$f_q(\zeta_q)= \frac{(1 + q)}{(q+2)^{1+q}} (q + 2- \zeta_q)^q $,
allowing for an easy interpolation between non-fluctuating ($q=-1$, $\zeta_{-1}=1$), 
uniform Dirac distributions and distributions increasingly skewed towards small 
$\zeta_q < 1$ for $q>-1$.

The jet-medium coupling is  $\kappa(T)=C_r\kappa^\prime(T)$ for quark ($C_r=1$)
and gluon ($C_r= \frac{C_A}{C_F}=\frac{9}{4}$) jets. Those jets are distributed 
according to a transverse initial profile specified by 
the bulk QGP flow fields given by the transverse plus Bjorken (2+1)d expansion 
of VISH2+1 \cite{VISH2+1}.

While this model can in principal be used to study a string-like jet-energy
loss \cite{Betz:2014cza,Betz:2012qq}, we strictly limit our discussion here to 
a radiative pQCD scenario including jet-energy loss fluctuations specified by 
$(a=0,z=1,c=3,q=0)$. This setting describes the data measured at
RHIC and LHC very well \cite{Betz:2014cza}.

\subsection{The Hybrid Strong/Weak Approach}
The jet-energy loss of the "hybrid strong/weak approach" \cite{Casalderrey-Solana:2014bpa}
is based on the falling string prescription by Chesler et al.\ \cite{Chesler:2014jva}
\begin{eqnarray}
\frac{1}{E_{\rm in}}\frac{dE}{dx} = -\frac{4}{\pi}\frac{x^2}{x^2_{\rm stop}}
\frac{1}{\sqrt{x^2_{\rm stop}-x^2}}\,,
\label{Eq_AdS}
\end{eqnarray}
featuring a Bragg peak. Here, the initial jet energy is $E_{in}$. 
The string stopping distance for quark and gluon jets is given by
\begin{eqnarray}
x^{q,g}_{\rm stop}=\frac{1}{2\kappa_{\rm sc}^{(g)}}\frac{E_{\rm in}^{1/3}}{T^{4/3}}
\end{eqnarray}
and the jet-medium coupling for gluon jets reads $\kappa_{\rm sc}^{(g)}=\kappa_{\rm sc}\left(
\frac{C_A}{C_F}\right)^{1/3}$, including the Casimir operators $C_A$ and $C_F$. 
This energy-loss ansatz is integrated into our existing mod\-el \cite{Betz:2014cza,Betz:2012qq}.
As for the pQCD scenario, we consider the (2+1)d viscous hydrodynamic bulk evolution of VISH2+1 
\cite{VISH2+1} as background, averaging over all possible initial energies $E_{in}$.

\begin{figure*}
\hspace*{3cm}
\includegraphics[scale=0.5]{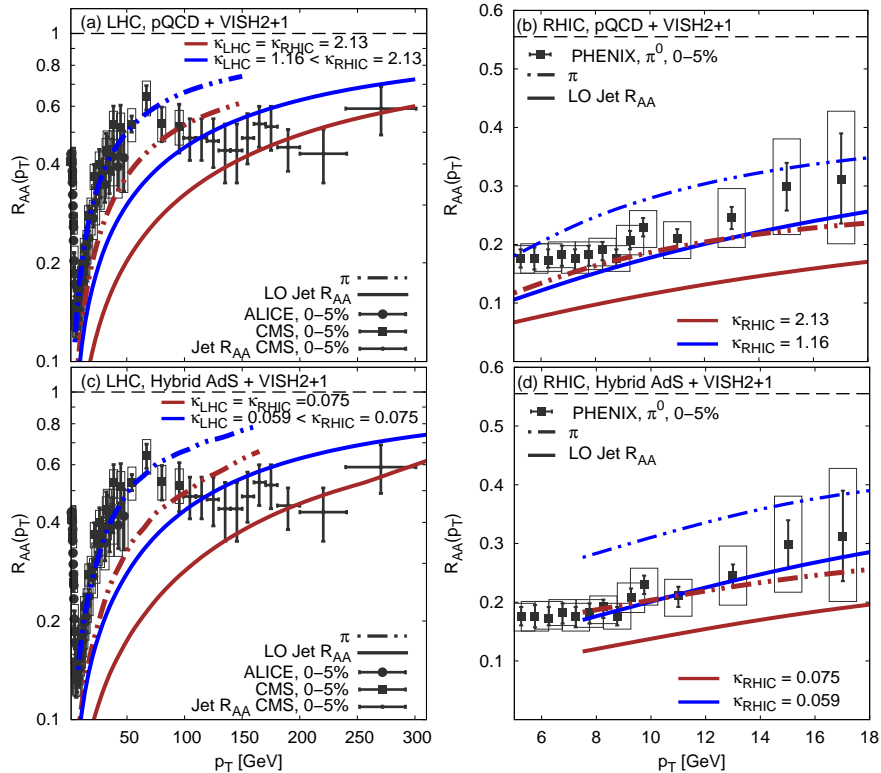}
\caption{{(Color online) The pion nuclear modification factor (dashed-dotted lines)
and the LO Jet $R_{AA}$ (solid lines) calculated via the radiative pQCD-like
energy-loss ansatz given by Eq.\ (\ref{Eq_pQCD}) with $(a=0,z=1,c=3,q=0)$ (upper panel)
and the hybrid strong/weak string energy-loss ansatz given by Eq.\ (\ref{Eq_AdS}) (lower panel) 
at LHC (left) and RHIC (right) energies for larger (red) and lower (blue) 
jet-medium couplings compared to the measured data \protect{\cite{Adare:2012wg,Abelev:2012hxa,CMS:2012aa,CMS:2012rba}}.}}
\label{fig1}       
\end{figure*}

\subsection{Model comparison}

In Ref.\ \cite{Betz:2014cza}, we considered an AdS/CFT inspired scenario
with $\frac{dE}{d\tau}=-\kappa(T)  E^0(\tau) \, \tau^{2} \, T^{4} \, \zeta_q$. 
The main difference between this ansatz and Eq.\ (\ref{Eq_AdS}) is the square-root 
dependence describing the Bragg peak with the explosive burst of energy close 
to the end of the jet's evolution.

There have been extensive discussions in literature \cite{Casalderrey-Solana:2014bpa,Ficnar,Betz:2008ka} 
on the impact of the Bragg peak, however, we are going to show below that
the difference of an AdS energy-loss model featuring a Bragg peak to a pQCD
model without a Bragg peak \cite{Betz:2014cza,Xu:2014ica,WHDG11,Renk:2011aa,Marquet:2009eq,Betz:2012qq,Betz:2008ka} is only mar\-gin\-al.

\subsection{Pion and Jet $R_{AA}$}

As in Refs.\ \cite{Betz:2014cza,Betz:2012qq} we use the KKP pion fragmentation 
functions \cite{Kniehl:2000hk} that have successfully been tested on the 
$pp\rightarrow\pi^0$ spectra at RHIC and LHC \cite{Simon:2006xt} to determine the
pion nuclear modification factors. 

In contrast to Ref.\ \cite{Casalderrey-Solana:2014bpa}, however, we do not reconstruct
jets with FASTJet \cite{FASTJet} but use the ansatz that
\begin{eqnarray}
{\rm Jet}\; R_{AA} = \frac{R_{AA}^g\,d\sigma_g(p_T)+R_{AA}^q\,d\sigma_q(p_T)}{d\sigma_g(p_T)+d\sigma_q(p_T)}
\end{eqnarray}
and the pQCD cross-sections from WHDG \cite{DGLV} to obtain
an idealized LO Jet $R_{AA}$. Naturally, this LO Jet $R_{AA}$ is only a lower bound
for the NLO Jet $R_{AA}$ with jet-cone radii $R>0$. For a detailed discussion
on the NLO and medium effects, please cf.\ to Ref.\ \cite{He:2011pd}.

\section{Results of the model comparison}
\label{Section3}
Figure \ref{fig1} shows the nuclear modification factor for pions (dashed-dotted lines)
and for the Jet $R_{AA}$ (solid lines) calculated via the pQCD jet-energy loss
ansatz of Eq.\ (\ref{Eq_pQCD}) (upper panel) and the hybrid strong/weak string 
energy loss given by Eq.\ (\ref{Eq_AdS}) (lower panel) at LHC (left) and RHIC (right) 
energies for two different jet-medium couplings, a larger one (red) and a lower one (blue),
compared to the data from PHENIX, ALICE, and CMS
\cite{Adare:2012wg,Abelev:2012hxa,CMS:2012aa,CMS:2012rba}.

The solid blue lines for the Jet $R_{AA}$ in the left panels of Fig.\ \ref{fig1} 
describe the experimental data extracted for the Jet $R_{AA}$ at LHC energies 
within the present error bars for a jet-medium coupling of $\kappa=1.16$ (pQCD, upper panel) 
and $\kappa=0.059$ (Hybrid AdS, lower panel), respectively. Fragmenting this result to obtain 
the pion nuclear modification factor (dashed-dotted lines) leads to an $R_{AA}$ 
that reproduces the measured pion nuclear modification factor at LHC as well. 
Please note that the yield of this pion nuclear modification factor is enhanced 
over the Jet $R_{AA}$ for the same jet-medium coupling which will be discussed below.

A straight extrapolation of this results to RHIC energies (right panels of Fig.\
\ref{fig1}), however, shows that the Jet $R_{AA}$ for {\it the same}
jet medium couplings of $\kappa=1.16$ (pQCD) or $\kappa=0.059$ (Hybrid AdS) (solid blue lines) 
by pure chance lie on top of the measured {\it pion} nuclear modification factor. 
Fragmenting this result to pions leads to a $R^{\pi}_{AA}$ that is larger than 
the measured data at RHIC.

For larger jet-medium couplings of  $\kappa=2.13$ (pQCD) 
and $\kappa=0.075$ (Hybrid AdS) (red lines), however, the {\it pion} nuclear
modification factor at RHIC is described (dashed-dotted red lines) but 
underpredicts the pion nuclear modification factor at the LHC, as known from
the "surprising transparency" at the LHC \cite{WHDG11}. On the other hand, the 
Jet $R_{AA}$ for this scenario also describes the extracted
data from ALICE and CMS within the present error bars.

Of course, this idealized LO Jet $R_{AA}$ is only a lower bound for the reconstructed 
nuclear modification factor and thus a Jet $R_{AA}$ extracted from reconstruction
algorithms will show a larger yield.

Please note that for this comparison, the jet-medium coupling has been treated 
as a constant. However, most recently \cite{Xu:2014ica} it has been shown that
the jet-medium coupling is not a constant but depends on the energy of the jet and the 
temperature of the medium.

Thus, in the large-$p_T$ range there are several competing effects:
\begin{enumerate}
\item a jet-medium coupling depending on the jet-energy and the temperature of the medium,
\item NLO and medium effects \cite{He:2011pd}, and
\item the jet-cone size $R$ enhancing the $R_{AA}$ for larger $R$.
\end{enumerate}
The extracted data suggest that these effects act in opposite directions tending to cancel out at the end
and leading to a $R_{AA}^\pi\sim {\rm Jet} R_{AA}$,
limiting the information of the reconstructed jets regarding the jet-medium coupling.

Besides that, Fig.\ \ref{fig1} shows that the results for the pQCD and the Hybrid
AdS energy-loss including a Bragg peak are remarkably similar. Thus, neither the 
pion nor the Jet $R_{AA}$ are sensitive to the difference in the path-length 
between pQCD and AdS models.

\section{Summary}
\label{Summary}

We compared the pion nuclear modification factor and a parton-jet nuclear modification factor describing the
LO Jet $R_{AA}$ at RHIC and LHC energies for a jet-energy loss based on radiative pQCD \cite{Betz:2014cza,Betz:2012qq} 
and on a hybrid strong/weak approach for falling strings \cite{Casalderrey-Solana:2014bpa}.
We found that for both scenarios the measured pion nuclear modification factor can only be described
considering a reduced jet-medium coupling at the LHC, while in case of the Jet $R_{AA}$ 
the experimental data can be characterized with the same jet-medium coupling at RHIC and LHC
as discussed in Ref.\ \cite{Casalderrey-Solana:2014bpa}.

For the Jet $R_{AA}$, however, there are several different competing effects: 
(1) a jet-medium coupling depending on the jet energy and the temperature of the medium \cite{Xu:2014ica},
(2) NLO and medium effects \cite{He:2011pd}, as well as (3) the impact of the jet cone radii $R$ increasing the 
Jet $R_{AA}$ for larger values of $R$ \cite{He:2011pd}. 
These effects tend to cancel each other, resulting in a $R_{AA}^\pi \sim {\rm Jet} R_{AA}$.

Altogether, these findings confirm our results of Ref.\ \cite{Betz:2014cza}.
Both, a pQCD-based jet-energy loss and a non-con\-for\-mal AdS/CFT-based approach
can describe the present experimental data. Besides that, the results based
on the pQCD and the AdS approach are insensitive to the jet-path length dependence 
of the models considered. Consequently, the path-length dependence 
for the jet-energy loss in heavy-ion collisions cannot be constrained further than $z=[0,2]$.

We conclude that while reconstructed jets are sensitive to both quarks and gluons and thus 
provide more information than the pion nuclear modification factor, their information
on the jet-medium coupling is limited due to the superposition with NLO and medium effects 
as well as the impact of the jet cone radii. Therefore, a complete description of the 
underlying jet physics requires to study both the leading hadron nuclear modification 
factor and the reconstructed jet nuclear modification factor.

\section*{Acknowledgments}
We are very grateful to U.\ Heinz and C.\ Shen for making 
their hydrodynamic field grids available and thank J.\ Xu and A.\ Ficnar for helpful
discussions. BB acknowledges financial support received from the 
Helmholtz International Centre for FAIR within the framework of the LOEWE 
program (Landes\-offensive zur Entwicklung Wis\-sen\-schaft\-lich-\"Okonomischer 
Exzellenz) launched by the State of Hesse. This work was supported in part from 
the US-DOE Nuclear Science Grant No.\ DE-FG02-93ER40764 and No.\ DE-AC02-05CH11231 
within the framework of the JET Topical Collaboration \cite{JET}.

\end{document}